\documentclass[a4paper,11pt]{article}

\pdfoutput=1
\usepackage{jcappub}

\usepackage{graphicx}
\usepackage{longtable}
\usepackage{float}
\usepackage{dcolumn}
\usepackage{bm}
\usepackage{appendix}
\usepackage{multirow}
\usepackage{color}
\usepackage[utf8]{inputenc}
\usepackage{footmisc}
\usepackage{hyperref}

\usepackage{aas_macros}
\usepackage[nomarkers,figuresonly,nolists]{endfloat}

\usepackage{xcolor}
\usepackage{graphicx}
\usepackage{bm}
\usepackage{ulem}
\usepackage{multirow, array, float, tabularx}
\usepackage{amsmath}
\hypersetup{
	citecolor = red,
	linkcolor = blue,
	urlcolor = magenta
}

\newcommand{\be}{\begin{equation}}
	\newcommand{\ee}{\end{equation}}
\newcommand{\bea}{\begin{eqnarray}}
	\newcommand{\eea}{\end{eqnarray}}
\newcommand{\bes}{\begin{subequations}}
	\newcommand{\ees}{\end{subequations}}

\newcommand{\bc}{\begin{center}}
	\newcommand{\ec}{\end{center}}

\usepackage{wasysym}

\begin{document}


\title{\boldmath J-PAS: Forecasting constraints on Neutrino Masses}



\author[a,b]{Gabriel Rodrigues,}
\author[b]{Antonio J. Cuesta,}
\author[a]{Jailson Alcaniz,}
\author[c]{Miguel Aparicio Resco,}
\author[d]{Antonio L. Maroto,}
\author[e]{Manuel Masip,}
\author[a]{Jamerson G. Rodrigues,}
\author[a]{Felipe B. M. dos Santos,}
\author[b]{Javier de Cruz P\'erez,}
\author[b]{Jorge Enrique Garc{\'\i}a-Farieta,}
\author[a]{Clarissa Siqueira,}
\author[f,g]{Fuxing Qin,}
\author[f,h]{Yuting Wang,}
\author[f,g,h]{Gong-Bo Zhao,}
\author[i,j]{Carlos Hernández–Monteagudo,}
\author[k,l,m]{Valerio Marra,}
\author[n]{Raul Abramo,}
\author[o]{Narciso Benítez,}
\author[p,q]{Silvia Bonoli,}
\author[a]{Saulo Carneiro,}  
\author[q]{Javier Cenarro,} 
\author[q]{David Cristóbal-Hornillos,}
\author[a]{Renato Dupke,}
\author[q]{Alessandro Ederoclite,} 
\author[q]{Antonio Hernán-Caballero,}
\author[q]{Carlos López-Sanjuan,}
\author[q]{Antonio Marín-Franch,}
\author[r]{Claudia Mendes de Oliveira,}
\author[q]{Mariano Moles,}
\author[r]{Laerte Sodré Jr.,}  
\author[s]{Keith Taylor,}
\author[q]{Jesús Varela,} 
\author[q]{and Héctor Vázquez Ramió} 

\emailAdd{gabrielrodrigues@on.br}
\emailAdd{ajcuesta@uco.es}
\emailAdd{alcaniz@on.br}
\emailAdd{miguel.aparicio@universidadeuropea.es}
\emailAdd{maroto@ucm.es}
\emailAdd{masip@ugr.es}
\emailAdd{jamersonrodrigues@on.br}
\emailAdd{fbmsantos@on.br}
\emailAdd{jdecruz@uco.es}
\emailAdd{jorge.farieta@uco.es}
\emailAdd{csiqueira@on.br}
\emailAdd{qinfx@nao.cas.cn}
\emailAdd{ytwang@nao.cas.cn}
\emailAdd{gbzhao@nao.cas.cn}
\emailAdd{chm@iac.es}
\emailAdd{valerio.marra@me.com}
\emailAdd{abramo@fma.if.usp.br}
\emailAdd{txitxo.benitez@gmail.com}
\emailAdd{silvia.bonoli@dipc.org}
\emailAdd{saulocarneiro@on.br}
\emailAdd{cenarro@cefca.es}
\emailAdd{dchornillos@gmail.com}
\emailAdd{rdupke@gmail.com}
\emailAdd{aederocl.astro@gmail.com}
\emailAdd{ahernan@cefca.es}
\emailAdd{clsj@cefca.es}
\emailAdd{amarin@cefca.es}
\emailAdd{claudia.oliveira@iag.usp.br}
\emailAdd{moles@cefca.es}
\emailAdd{laerte.sodre@iag.usp.br}
\emailAdd{kt.astro@gmail.com}
\emailAdd{jvarela@cefca.es}
\emailAdd{hvr@cefca.es}

\affiliation[a]{Observatório Nacional, Rua General José Cristino, 77, São Cristóvão, 20921-400, Rio de Janeiro, RJ, Brazil}

\affiliation[b]{Departamento de F{\'\i}sica, Universidad de C\'ordoba,
Campus Universitario de Rabanales, Ctra. N-IV Km.~396, E-14071 C\'ordoba, Spain}

\affiliation[c]{Universidad Europea de Madrid. Campus de Villaviciosa de Od\'on. C/ Tajo, s/n. Urb. El Bosque, Villaviciosa de Od\'on, 28670 Madrid, Spain}

\affiliation[d]{Departamento de F{\'\i}sica Te\'orica and Instituto de F{\'\i}sica de Part{\'\i}culas y del Cosmos (IPARCOS-UCM),
Universidad Complutense de Madrid, 28040 Madrid, Spain}

\affiliation[e]{CAFPE and Departamento de F{\'i}sica Te\'orica y del Cosmos, Universidad de Granada, E-18071 Granada, Spain}

\affiliation[f]{National Astronomical Observatories, Chinese Academy of Sciences, 100101 Beijing, P.R.China}

\affiliation[g]{School of Astronomy and Space Sciences, University of Chinese Academy of Sciences, 100049 Beijing, P.R.China}

\affiliation[h]{Institute for Frontiers in Astronomy and Astrophysics, Beijing Normal University, 102206 Beijing, China}

\affiliation[i]{Instituto de Astrofísica de Canarias, C/ Vía Láctea, s/n, E-38205, La Laguna, Tenerife, Spain}

\affiliation[j]{Universidad de La Laguna, Avda. Francisco Sánchez, E-38206, San Cristóbal de La Laguna, Tenerife, Spain}

\affiliation[k]{Departamento de Física, Universidade Federal do Espírito Santo, 29075-910, Vitória, ES, Brazil}

\affiliation[l]{Osservatorio Astronomico di Trieste, via Tiepolo 11, 34131 Trieste, Italy (INAF)}

\affiliation[m]{Institute for Fundamental Physics of the Universe, via Beirut 2, 34151, Trieste, Italy (IFPU)}

\affiliation[n]{Departamento de Física Matemática, Instituto de Física, Universidade de São Paulo, Rua do Matão 1371, 05508-090, São Paulo, SP, Brazil}

\affiliation[o]{Independent Researcher}

\affiliation[p]{Donostia International Physics Center (DIPC), Manuel Lardizabal Ibilbidea, 4, San Sebastián, Spain}

\affiliation[q]{Centro de Estudios de Física del Cosmos de Aragón (CEFCA), Plaza San Juan, 1, E-44001, Teruel, Spain}

\affiliation[r]{Departamento de Astronomia, Instituto de Astronomia, Geofísica e Ciências Atmosféricas, Universidade de São Paulo, São Paulo, Brazil}

\affiliation[s]{Instruments4, 4121 Pembury Place, La Canada Flintridge, CA 91011, U.S.A.}

\abstract{The large-scale structure survey J-PAS is taking data since October 2023. In
this work, we present a forecast based on the Fisher matrix method to establish its sensitivity to the sum of the neutrino masses. We adapt the Fisher Galaxy Survey Code (FARO) to account for the neutrino mass under various configurations applied to galaxy clustering measurements. This approach allows us to test the sensitivity of J-PAS to the neutrino mass across different tracers, with and without non-linear corrections, and under varying sky coverage. We perform our forecast for two cosmological models: $\Lambda CDM + \sum m_\nu$ and $w_0w_a CDM + \sum m_\nu$. We combine our J-PAS forecast with Cosmic Microwave Background (CMB) data from the Planck Collaboration and Type Ia supernova (SN) data from Pantheon Plus. Our analysis shows that, for a sky coverage of 8,500 square degrees, J-PAS galaxy clustering data alone will constrain the sum of the neutrino masses to an upper limit at 95\% C.L of $\sum m_\nu < 0.32$ eV for the $\Lambda CDM + \sum m_\nu$ model, and $\sum m_\nu < 0.36$ eV for the $w_0w_a CDM + \sum m_\nu$ model. When combined with Planck data, the upper limit improves significantly. For J-PAS+Planck at 95\% C.L, we find $\sum m_\nu < 0.061$ eV for the $\Lambda CDM + \sum m_\nu$ model, and for J-PAS+Planck+Pantheon Plus, we obtain $\sum m_\nu < 0.12$ eV for the $w_0w_a CDM + \sum m_\nu$ model. These results demonstrate that J-PAS clustering measurements can play a crucial role in addressing challenges in the neutrino sector, including potential tensions between cosmological and terrestrial measurements of the neutrino mass, as well as in determining the mass ordering.
}

\vskip 1.0cm



\maketitle

\vskip 1.0cm

\section{Introduction}

The mechanism behind the origin of particle masses was revealed after the award-winning work of Englert and Higgs~\cite{Englert:1964et, Higgs:1964pj,Higgs:1964ia}. The key ingredient in this mechanism is the Higgs field, a scalar field with a non-zero vacuum expectation value that provides a mass term to the particles in the standard model of particle physics (SM). This mechanism was supported by the discovery of a Higgs-like boson at the Large Hadron Collider (LHC) in 2012~\cite{aad21, Chatrchyan12}. Neutrinos, however, are the exception. Only left-handed (active) neutrinos participate in weak interactions, resulting in massless neutrinos for the SM particle content. The appearance of neutrino masses requires the addition of new fields whose mass is not protected by the SM symmetry, or of higher-dimensional operators that break lepton number \cite{Weinberg:1979sa}. Therefore, the mechanisms usually used to describe the smallness of neutrino masses require new particles above the electroweak scale.

The scale of neutrino masses was apparent in the anomalies found in solar and atmospheric neutrinos, which point to an oscillatory behavior between the three flavors explained by the nondiagonal structure of their mass matrix. In particular, the set of parameters~\cite{Esteban:2020cvm,salas2021,ParticleDataGroup:2024cfk},
\begin{eqnarray}
   &&\sin^2{\theta_{12}} = 3.07^{+0.12}_{-0.11}\,\times \, 10^{-1}, \,\,\, \sin^2{\theta_{23}} = 5.61^{+0.12}_{-0.15}\,\times \, 10^{-1},\nonumber \\
    &&\sin^2{\theta_{13}} = 2.195^{+0.054}_{-0.058}\,\times \, 10^{-2},\nonumber \\
    &&\Delta m^2_{21} = 7.49^{+0.19}_{-0.19}\, \times \, 10^{-5}, \,\, \,\,\Delta m^2_{32} = 2.534^{+0.025}_{-0.023}\, \times \, 10^{-3} \label{Eq:Osc}
\end{eqnarray}
with the quadratic mass differences $m^2_{ij} = m^2_i - m^2_j$ evaluated in eV$^2$, provide the best fit for the data. A lower limit for the sum of neutrino masses is then obtained from this information. For the Normal Hierarchy spectrum (NH), $m_1 < m_2 \ll m_3$, one obtains $\sum m_i \gtrsim 0.057$\;eV, while for the Inverted Hierarchical scheme (IH), $m_3 \ll m_1 < m_2$, one may infer $\sum m_i \gtrsim 0.099$\;eV, where the sum runs over the different mass states. 

On the other hand, laboratory experiments such as KATRIN \cite{KATRIN:2001ttj} (tritium beta decays) are rapidly improving the constraints on the absolute value of the effective electron neutrino mass, placing an upper limit of 0.45\;eV on the effective electron anti-neutrino mass at the 90\% confidence level \cite{Katrin:2024tvg}. Combined with searches for neutrinoless double beta decays \cite{GERDA:2020xhi}, they could establish the absolute value and the Dirac or Majorana nature of this mass in the near future.

Complementary to oscillation and laboratory experiments, cosmological observations may place stringent constraints on neutrino masses. The presence of a light, stable, weakly interacting particle in the cosmological fluid can affect its evolution, both at the background and perturbative level. The main effect of the neutrinos on the evolution of cosmological perturbations is dictated by their free-streaming, which acts to dampen the metric perturbations at small length scales. For neutrinos transitioning to the non-relativistic regime in a matter-dominated universe, the comoving wavenumber associated with the free-streaming scale reaches a minimum at about $k_{nr} \simeq 0.018\, \Omega_m^{1/2}(\sum m_\nu/\text{eV})^{1/2}\, h$ Mpc$^{-1}$ \cite{Lesgourgues:2012uu,Lesgourgues:2013sjj}, setting the scale above which perturbations are suppressed. 

The resulting scale-dependent suppression of the matter power spectrum makes cosmological surveys particularly useful in constraining the sum of the neutrino masses (see e.g. \cite{Cuesta:2015iho}). This is especially true when early- and late-time probes of cosmological evolution, such as Cosmic Microwave Background (CMB) data and galaxy distribution observations, are jointly analyzed. In particular, the angular power spectrum of Planck temperature and polarization data, including CMB lensing, constrains the total neutrino mass to 0.24~eV \cite{Planck:2018vyg}. Consequently, a wide variety of cosmological data have recently been used to set an upper limit on the sum of the neutrino masses \cite{Palanque-Delabrouille:2019iyz,DiValentino:2021hoh,DiValentino:2022njd}, with potential implications for defining their mass hierarchy.


Recent analysis of the second-year data release from the Dark Energy Spectroscopic Instrument (DESI) \cite{desi16}, combined with cosmic microwave background (CMB) temperature, polarization and (Planck and ACT) lensing information, indicates a preference for the normal hierarchy (NH) spectrum over the inverted hierarchy, reducing the $ 95\%$ confidence limit on neutrino masses to $\sum m_\nu < 0.064$~eV, and showing a preference for a zero to negative values~\cite{DESI25}. The exact value of these bounds is model-dependent, with some of the most restrictive results arising from the standard cosmological model $\Lambda CDM +\sum m_\nu$. On the other hand, dynamical dark energy (DDE) models, specifically those based on phenomenological parameterizations of the dark energy equation of state (EoS), such as the Chevallier–Polarski–Linder (CPL) model, relax this constraint to the point where there is no tension, and both neutrino mass hierarchies are consistent. More recently, physically motivated DDE models, such as thawing quintessence, have been shown to constrain the total neutrino mass at the same level as the standard model $\sum m_\nu < 0.07$~eV~\cite{Gabriel25}.

Nevertheless, there is a clear trend toward decreasing the upper limit on neutrino masses as new datasets are considered. The authors in \cite{Jiang:2024viw}, by assuming an $\Lambda CDM +\sum m_\nu$ model, placed the upper limit at $\sum m_\nu < 0.042$~eV considering the same CMB data together with DESI DR1, galaxy cluster angular diameter distance observations, and a $H_0$ prior from SH0ES. The implications of this latest analysis extend beyond the hierarchy discussion (see, e.g. \cite{Jimenez:2022dkn,gariazo2022}) and highlight the potential for an emerging tension between cosmological observations and oscillation data. It is imperative to obtain new and accurate data to either dismiss or confirm this tension.

The Javalambre-Physics of the Accelerated Universe Astrophysical Survey (J-PAS) \cite{Benitez14} is currently collecting positions and spectrophotometric information of galaxies up to redshift $z\lesssim 1.0$. The miniJPAS survey \cite{Bonoli21}, which serves as a proof of concept for the full survey, using the J-PAS Pathfinder camera and the 56 J-PAS filters (54 narrow-band, FWHM 145\AA, and two broader filters extending to the UV and the near-infrared, complemented by the $u,g,r,i$ SDSS broad-band filters) already carried out a $\sim 1$\;deg$^2$ survey on the AEGIS field (along the Extended Groth Strip).

This work presents a standalone forecast for the J-PAS galaxy clustering probe, serving as a baseline quantification of its intrinsic constraining power on the sum of neutrino masses. We explore J-PAS's potential for cosmological parameter estimation, particularly its sensitivity to the sum of neutrino masses, within the standard $\Lambda$CDM as well as the dynamical dark energy framework. These Fisher forecasts are important for assessing the survey's expected constraining power and scientific return. 
The paper is organized as follows. In Section 2, we describe the methodology and the survey specifications considered to forecast the results. In Section 3, we present the results, both for the forecasts alone and for the combination of J-PAS with other datasets. In Section 4, we summarize our findings and discuss future perspectives.

\section{Methodology}
\subsection{A review on the Fisher matrix formalism}

One of the most straightforward ways to forecast future constraints of an experiment is the Fisher matrix formalism \cite{Tegmark:1996bz,Tegmark:1997rp,Coe:2009xf,Amendola:2015ksp}. The main advantage of this approach is that we do not need to make mock data, so it is a fast formalism.

Given a likelihood \(\mathcal{L}(\mathbf{x}|\boldsymbol{\theta})\) for data vector \(\mathbf{x}\) and model parameters \(\boldsymbol{\theta}\), the Fisher matrix is defined as the expectation value of the curvature of the log-likelihood,
\begin{equation}\label{eq:F_general}
F_{\alpha\beta}
\equiv -\left\langle
\frac{\partial^2\ln\mathcal{L}}{\partial\theta_\alpha\partial\theta_\beta}
\right\rangle.
\end{equation}
Assuming a Gaussian distribution for the data, the likelihood can be written as
\begin{equation}\label{LH}
{\cal L} \propto e^{-\frac{\chi^2}{2}} \equiv \mathrm{exp} \left[ -\frac{1}{2} \sum_{i \, j} \left(x_i - \tilde{x}_i \right) C^{-1}_{ij} \left(x_j - \tilde{x}_j \right) \right],
\end{equation}
where $x_i$ are the observables, $\tilde{x}_i$ is the prediction of a model with parameters $\theta_k$ and $C_{ij}$ is the covariance matrix of the data.
Although the likelihood, defined as (\ref{LH}), is Gaussian in the observables $x_i$, the likelihood of the parameters $\theta_k$ is not necessarily Gaussian. The Fisher matrix formalism approximates the likelihood of the parameters $\theta_k$ as Gaussian around fiducial values $\tilde{\theta}_k$. Given this approximation, a linear change of variable can be made from the Fisher matrix of the data to the Fisher matrix of the parameters:
\begin{equation}
F_{kl}^{\theta} = \left.\frac{\partial x_i}{\partial \theta_k}\right|_{\tilde{\theta}} \, F_{ij}^{x} \, \left.\frac{\partial x_j}{\partial \theta_l}\right|_{\tilde{\theta}}.
\end{equation}
%
%

\subsection{The FARO code}

To perform this forecast, we use the Fisher gAlaxy suRvey cOde ($\texttt{FARO}$)\footnote{\url{https://www.ucm.es/iparcos/faro}} \cite{Resco21,AparicioResco:2019hgh}. $\texttt{FARO}$ is a Python code that computes the Fisher matrix for galaxy survey observables. The main approach of the code is to perform a tomographic analysis in a model-independent way. In the case of galaxy map observables, we will focus only on the multi-tracer galaxy distribution power spectra \cite{Abramo:2011ph,Abramo:2019ejj}.

The main parameters of the code for the multitracer power spectra are: $A_i (z) \equiv \sigma_{8}(z) \, b_i (z)$, $R(z) \equiv \sigma_{8}(z) \, f(z)$ and $E(z) \equiv H(z)/H_0$. Where $b_i(z)$ is the bias for the tracers, $H(z)$ is the Hubble parameter, and $\sigma_8(z) = \sigma_8 D(z)$, where $D(z)$ is the linear growth factor and $ \sigma_8 $ is the normalization of the matter power spectrum on scales of $ 8 \, h^{-1} $ Mpc today. The growth function $f(z)$ is defined as $D(z) = \exp \left(- \int_0^z \frac{f(z')}{1 + z'} dz' \right)$. It is indeed true that the linear growth factor $D$ and the growth rate $f$ acquire a weak scale dependence in the presence of massive neutrinos. However, as pointed out in section 2.5 of~\cite{Blanchard20}, this effect is only sub-percent. This approximation is fully adequate for the sole purpose of projecting the survey’s sensitivity. We note also that the linear growth factor can exhibit a stronger scale dependence at early times, while at the low redshifts relevant to J-PAS this dependence remains very weak. 

Also, a model-independent parametrization is performed for $P(k)$ as a function of $a_n$ parameters that modify the fiducial $P(k)$ in each different $k-$bin.

We have considered three effects for the power spectrum: the Kaiser effect for redshift space distortions \cite{Kaiser:1987qv}, the convolution of the redshift error term \cite{Amendola:2015ksp}, and the Alcock-Paczynski effect \cite{Alcock:1979mp}, with the multi-tracer power spectra given by,

\begin{equation}
P_{ab}^{\delta\delta}(z, \hat{\mu}_r, k_r) =\frac{D_{A,r}^2 E}{D_A^2 E_r}   (A_a + R \hat{\mu}^2) (A_b + R \hat{\mu}^2) \hat{P}(k) 
\exp\left(-k_r^2 \hat{\mu}_r^2 \frac{\sigma_a^2}{2}\right)
\exp\left(-k_r^2 \hat{\mu}_r^2 \frac{\sigma_b^2}{2}\right).
\end{equation}
Here, the subscript $r$ indicates that the corresponding quantity is evaluated in the fiducial model, and $\hat{\mu} $ is the angle between the wavevector $\mathbf{k}$ and the line of sight. 
$D_A(z)$ is the angular diameter distance for a flat universe, given by $D_A(z) = (1 + z)^{-1} \chi(z)$, where,

\begin{equation}
    \chi(z) = H_0^{-1} \int_{0}^{z} \frac{dz'}{E(z')}.
\end{equation}

The term $\sigma_{ab} = \frac{\delta z_{ab}^C (1 + z)}{H(z)}$ represents the radial error for the tracers, where the redshift error in the clustering measurements is defined as $\delta z_{ab}^C (1 + z)$. The Fisher matrix for the 3D multitracer power spectrum is given by:
\[
\begin{aligned}
F_{\alpha\beta}^{\delta\delta} =
\sum_{i,c,m} V_i & \, \Delta \hat{\mu}_m \, \Delta \log k_c \, k_c^3 \, \frac{3c}{8\pi^2}
\frac{\partial P_{ab}^{\delta\delta}(z_i, \hat{\mu}_m, k_c)}{\partial p_\alpha} \,
C^{-1}_{ba'} \\
& \times \frac{\partial P_{a'b'}^{\delta\delta}(z_i, \hat{\mu}_m, k_c)}{\partial p_\beta} \,
C^{-1}_{b'a} \,
\exp\left(-k_c^2 \Sigma_\perp^2 - k_c^2 \hat{\mu}_m^2 (\Sigma_\parallel^2 - \Sigma_\perp^2)\right),
\end{aligned}
\]

The integrals in $\hat{\mu}$ have an interval from $ -1$ to $1$, and $k$ ranges from $k_{\text{min}}$ to $\infty$. The code sets the value $k_{\text{min}} = 7\times10^{-3} \, h/\text{Mpc}$. The quantities $\Sigma_\perp(z) = 0.785 \, D(z) \, \Sigma_0$ and $\Sigma_\parallel(z) = 0.785 \, D(z) \, (1 + f(z)) \, \Sigma_0$ represent the exponential cut-off, which removes the contribution from non-linear scales~\cite{Seo07}. The covariance matrix $C_{ab}$ is given by

\begin{equation}
    C_{ab}(z_i, \hat{\mu}_m, k_c) = P_{ab}^{\delta\delta}(z_i, \hat{\mu}_m, k_c) + \frac{\delta_{ab} }{\bar{n}_a(z_i)},
\end{equation}
where $\bar{n}_{ab}(z_i)$ is the galaxy density of the tracers in the bins of redshift $z$ and, $V_i$ is the total volume of the $ i$-th bin, given by,
\begin{equation}
    V_i = \frac{4 \pi f_{\text{sky}}}{3} \left[ \chi(\bar{z}_i)^3 - \chi(\bar{z}_{i-1})^3 \right],
\end{equation}
$f_{\text{sky}}$ is the sky fraction of the survey and $\bar{z}_i$ is the upper limit of the $i$-th bin. The Kronecker delta $\delta_{ab}$ ensures that shot noise contributes only to the auto-correlation of each tracer, effectively expressing the absence of shot noise in cross-correlations between different tracer types. The FARO code will then project the Fisher matrix for the observable power spectra into the parameters we want to constrain.

The standard FARO code also has the possibility to project the information of each tomographic parameter into another set of parameters like, for example, a simple $w_0w_a CDM$ dark energy model. Given the initial Fisher matrix for model-independent parameters $\{p_\alpha\}$, we can obtain the Fisher matrix for a new set of parameters $\{q_\alpha\}$ as follows:
\begin{equation}
\textbf{F}^{q}=\textbf{P}^{t} \, \textbf{F}^{p} \, \textbf{P},
\label{eq:change}
\end{equation}
where $P_{\alpha\beta}=\partial{p_{\alpha}}/\partial{q_{\beta}}$, evaluated on the fiducial model. Even though this approach is straightforward to implement for redshift dependent parameters, it can be somewhat complicated for the $P(k)$ model-independent parametrization. To solve this inconvenience, we have modified the code to ensure that this change of variables can be done in the simplest way. Instead of obtaining the Fisher matrix for tomographic parameters and then doing the projection, we implement the projection inside the code to directly obtain the Fisher matrix for the desired cosmological parameters. This allows us to calculate $\partial P(k) /\partial q_{\alpha}$ in a straightforward way, solving the issue of computing how the model-independent parametrization of $P(k)$ depends on the chosen cosmological parameters.  Most derivatives inside the code are done analytically, except for the derivatives of the matter power spectrum with respect to the cosmological parameters. These derivatives are computed numerically. We also assume that for dynamical dark energy models, $\partial \hat{P}/\partial w_0=\partial \hat{P}/\partial w_a=0$, where $\hat{P}\equiv P/\sigma_8^2$. 

In our template the effect of neutrino masses is constrained to the normalized power spectrum $\hat{P}(k)$. Besides, at the redshift values relevant for large scale structure observations, background effects due to non-zero $\Sigma m_\nu$ are already accounted for in $\Omega_m$. Therefore, the only explicit dependence of our model on the sum of neutrino masses is in the derivative $\partial\hat{P}/\partial\Sigma m_\nu$. Moreover, the fact that $\Sigma m_\nu$ is physically bounded from below to non-negative values is dealt with at the posterior level (i.e. not in the Fisher matrix), including a step function prior that removes negative values when necessary.

\subsection{Convergence and consistency tests}

We use the vibration matrix method \cite{Yahia-Cherif21} to check the stability of the Fisher matrix with respect to the individual numerical derivatives. In particular, we find that a 5-point stencil is stable with an epsilon value of $\varepsilon=10^{-2}$. This is the optimal choice, as it is the smallest possible value that does not introduce artificial numerical noise. In addition, the new version of $\texttt{FARO}$ was tested against the forecast for the $w_0w_a CDM$ model using EUCLID \cite{Casas24,Blanchard20} and obtained consistent results. For the EUCLID consistency check, we assume a sky area of $15000$\;deg$^2$, along with the number densities, redshift bins, and redshift errors shown in Tables I and II of~\cite{Resco21}. We have also compared the results from the Fisher matrix directly using the parameter space of the cosmological parameters, and indirectly using the model-independent parameter space of \texttt{FARO}, and then projecting this Fisher matrix on the cosmological parameter space, using the change of variable equation (Eq.~\ref{eq:change}). For our constraints, we have used a modified version of \texttt{FARO} (\texttt{FARO} 2.0) which works on the cosmological parameter space directly and also can run using \texttt{Python} 3.0.

A critical test here is that, to perform the change of variables, the model-independent parameters $E_i\equiv E(z_i)=H(z_i)/H_0$ have to be replaced by $H_i\equiv H(z_i)$ so that the numerical derivative $\partial E/\partial h$ was non-zero. The derivative $\partial P/\partial h$ was performed numerically to avoid using the chain rule given the $h$ dependence of $k$ (see discussion at Section 4.5.1 in \cite{Blanchard20}).

\subsection{Forecast settings}
To compute the derivatives and cosmological quantities that are used in the Fisher matrices, we need to choose fiducial values for the parameters. Here, we explore two different cosmological models, $\Lambda CDM+\sum m_\nu$ and $w_0w_a CDM + \sum m_\nu$. Thus, we set fiducial values for the Hubble constant $h$, the matter density parameter $\Omega_m$, the baryon density parameter $\Omega_b$, the spectral index $n_s$, the amplitude of perturbations $\sigma_8$, the total neutrino mass $\sum m_\nu$, and the dark energy equation of state parameters $w_0$ and $w_a$. Table~\ref{Tab1} presents the values that have been used. These are the fiducial values defined by the J-PAS collaboration, they are the same as those used by the Planck collaboration~\cite{planck13} and also used by the project Universe N-body simulations for the Investigation of Theoretical models from galaxy surveys (UNIT)~\cite{Angulo16}.

\begin{table}[ht]
\centering
\begin{tabular}{|c|c|} 
\hline
Parameter & Value                                                     \\ 
\hline
\hline
{\boldmath$h$}& $0.6774$    \\

{\boldmath$\Omega_m$} &  $0.3089$   \\
				
{\boldmath$\Omega_{b}$} & $0.0223$   \\
		
{\boldmath$n_s$}& $0.9667$  \\
{\boldmath$\sigma_{8}$}& $0.8161$  \\

{\boldmath$\sum m_\nu$ (eV)}& $0.06$  \\ 

{\boldmath$w_0$}& $-1$  \\

{\boldmath$w_a$}& $0.0$  \\ 
\hline
\hline

\end{tabular}
\caption{Fiducial values for the cosmological parameters.}
\label{Tab1}
\end{table}

In addition to the fiducial values, we need to define the survey specifications. Here, we consider two different tracers: Luminous Red Galaxies (LRGs) and Emission Line Galaxies (ELGs). The fiducial bias for these is given by:
\begin{equation}
    b(z) = \frac{b_0}{D(z)},
\end{equation}
where $D(z)$  is the growth function, with $b_0 = 0.84$ for ELGs and $b_0 = 1.7$ for LRGs~\cite{desi16}. In order to account for a possible residual redshift dependence, we have also tested a bias function with an additional degree of freedom, $b(z) = \frac{b_0 + c_0 z}{D(z)}$. In this case, $c_0$ is treated as a nuisance parameter, and we marginalize over it. As a result, we observed only minor changes in the overall neutrino mass constraints. Therefore, we used the simplest bias function as our fiducial model. We divided our data into four redshift bins, ranging from $z = 0.1$ to $z = 0.7$. In Table~\ref{tab:2}, we present the $b(z)$ values for each redshift bin and tracer. 

It should be noted that we do not use the galaxy number densities estimated pre-observations with JPCam mentioned in \cite{Benitez14}. Instead we use currently measured values based on real observations. For a similar forecast comparing values of galaxy number densities ''pre-observation'' J-PAS with PFS and future CMB experiments, see \cite{Qin:2025nkk}. In the forecasts presented in this work, we make use of galaxy number density estimates in redshifts and odds bins obtained from J-PAS Internal Data Release (IDR) of June 2024. 
In particular, we use about 27~sq.deg. with full filter coverage, and where the $i$-band is used as detection band under exposure times that are deemed as representative for the full J-PAS survey. This data set incorporates the latest calibration pipeline, including correction for the temporal instability of the bias. The photometric redshifts (photo-$z$s) have been obtained with the J-PAS implementation of {\tt LePhare} \cite{AntonioHC_miniJPASphotozs,Caballero_2023}. We tested three different star-galaxy separation codes, independently developed within the collaboration and since the differences among them were not significant, we quote results for the one that is based exclusively upon morphological information of the sources, together with their $i$-band magnitude. We also rely on a red/blue classification provided as in \cite{DiazGarcia_2024} in order to provide estimates of the number density of each type of galaxy for different redshift and odds bins. 


\begin{table*}[!h]
    \centering
    \begin{tabular}{|l|l|l|}
        \hline
        \hline
        z   & $b(z)$ for ELG & $b(z)$ for LRG \\
        \hline
        \hline
        0.1 & 0.888 & 1.797 \\
        0.3 & 0.986 & 1.996 \\
        0.5 & 1.0953 & 2.21 \\
        0.7 & 1.208 & 2.446 \\
        \hline

    \end{tabular}
   \caption{Bias function values for each redshift bin and tracer.}
    \label{tab:2}
\end{table*}

The probe considered in this work is galaxy clustering (GCs). The galaxy number densities and redshift errors were divided into three distinct subsamples based on their odds. The first subsample contains galaxies with the top 30\% odds, the second includes galaxies with odds in the range of 50\% to 70\%, and the third comprises galaxies with the lower 50\% odds. For blue galaxies, we used all available odds, while for red galaxies, we restricted our analysis to the top 30\%. This decision was made due to the low galaxy counts for LRGs in the remaining 20\% and the lower 50\% odds, which resulted in redshift errors that were too small to be reliable.

Our implementation of {\tt LePhare} also provides {\tt PHOTOZ-ERR} which is an estimate of the width of the photo-$z$ probability density distribution (PDF), the details of which can be seen in \cite{Caballero_2023}). Table~\ref{tab:3} and Table~\ref{tab:4} show the redshift bins, the galaxy number densities in units of $\:h^3 \: \text{Mpc}^{-3}$, and the redshift errors for the clustering case, for LRGs and ELGs, respectively.

\begin{table*}[!h]
    \centering
    \begin{tabular}{|l|l|l|}
        \hline
        \hline
        z   & nLRG $[h^3\;Mpc^{-3}]$ & $\delta z^c$LRG \\
        \hline
        \hline
        \multicolumn{3}{|c|}{\textbf{Top 30 \%}} \\
        \hline
        0.1 & 0.0036 & 0.0023 \\
        0.3 & 0.0027 & 0.0023 \\
        0.5 & 0.00072 & 0.0038 \\
        0.7 & 0.00037 & 0.0043 \\
        \hline

    \end{tabular}
    \caption{Redshift bins, number densities and redshift errors of luminous red galaxies (LRG) for J-PAS.}
    \label{tab:3}
\end{table*}

\begin{table*}[!h]
    \centering
    \begin{tabular}{|l|l|l|}
        \hline
        \hline
        z   & nELG $[h^3\;Mpc^{-3}]$ & $\delta z^c$ELG \\
        \hline
        \hline
        \multicolumn{3}{|c|}{\textbf{Top 30 \%}} \\
        \hline
        0.1 & 0.011 & 0.0019 \\
        0.3 & 0.0054 & 0.0019 \\
        0.5 & 0.0026 & 0.0058 \\
        0.7 & 0.0013 & 0.0054 \\
        \hline
        \multicolumn{3}{|c|}{\textbf{Remaining 20 \%}} \\
        \hline
        0.1 & 0.0086 & 0.0026 \\
        0.3 & 0.0047 & 0.0043 \\
        0.5 & 0.0022 & 0.011 \\
        0.7 & 0.0011 & 0.013 \\
        \hline
        \multicolumn{3}{|c|}{\textbf{Lower 50 \%}} \\
        \hline
        0.1 & 0.028 & 0.039 \\
        0.3 & 0.013 & 0.033 \\
        0.5 & 0.0057 & 0.039 \\
        0.7 & 0.0027 & 0.033 \\
        \hline
    \end{tabular}
    \caption{Redshift bins, number densities and redshift errors of emission line galaxies (ELG) for J-PAS.}
    \label{tab:4}
\end{table*}

\begin{figure}[ht]
    \centering
    \includegraphics[scale=0.60]{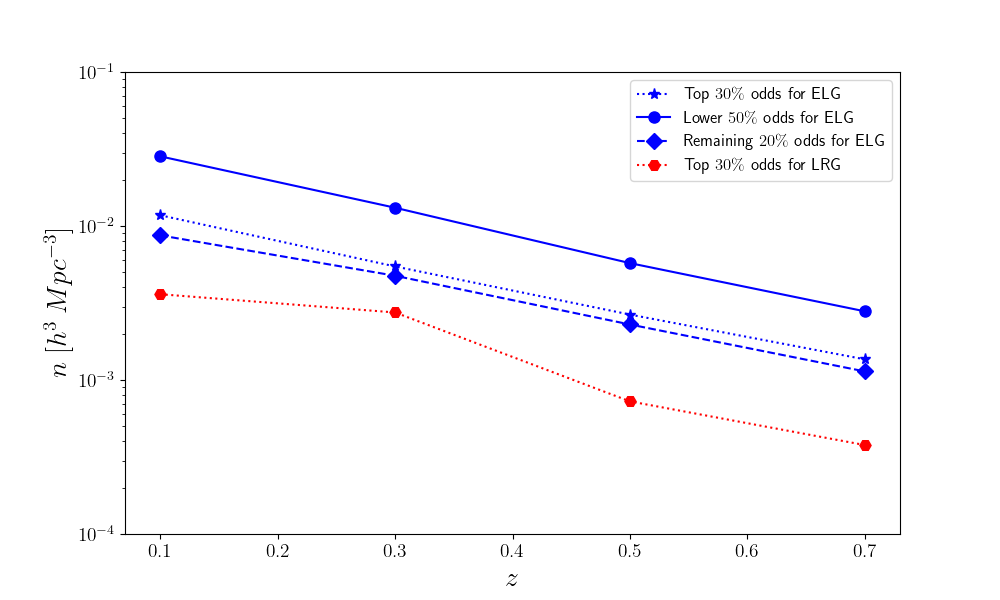} \quad
    \includegraphics[scale=0.60]{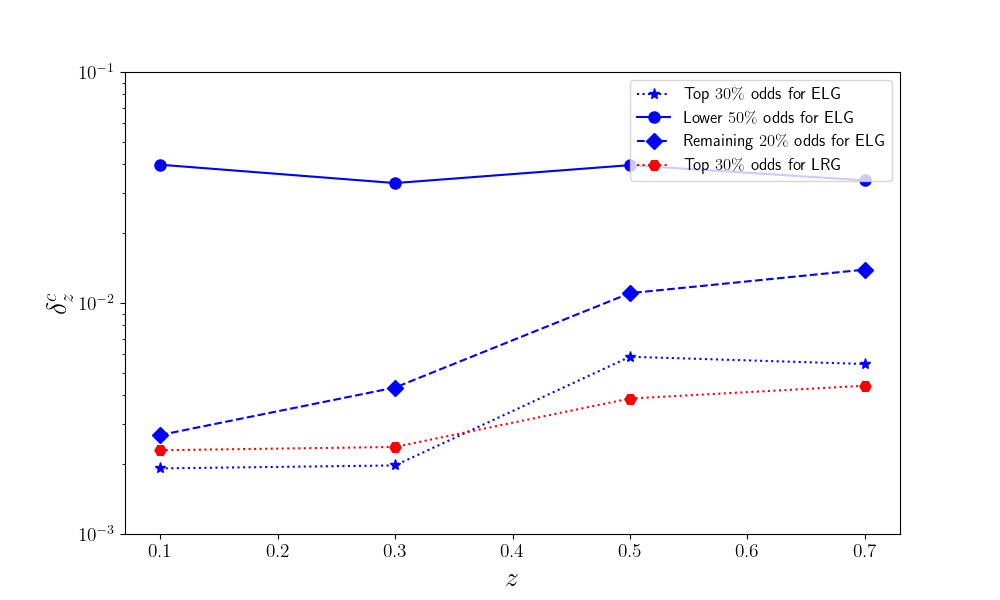}
    \caption{Galaxy number density in units of $\:h^3 \: \text{Mpc}^{-3}$ and redshift errors for each divided sub-sample and each tracer as a function of redshift.}
    \label{fig:1}
\end{figure}

We have explored two sky coverages here: $f_{\text{sky}} = 1500 \:\text{deg}^2$ for the forecast of constraints from the expected analysis of initial results, and $f_{\text{sky}} = 8500 \:\text{deg}^2$ for the forecast of the full J-PAS operation. It is important to mention that the speed of J-PAS sky coverage has not been defined yet.  In Fig.~\ref{fig:1}, we show the distribution of number densities for each sub-sample and each type of galaxy, as well as the redshift error for each redshift bin. 

Due to the suppression that massive neutrinos add to the power spectrum, which is larger on small scales (large $k$-values), we are going to model the matter power spectrum into the non-linear regime. For this analysis, we considered scales that range from $k=7\times 10^{-3} \, h/\text{Mpc}$ to $k_{\text{max}}$. To model the non-linear power spectrum, we have used the Cosmic Linear Anisotropy Solving System code (CLASS) \cite{lesgourgues11,Blas:2011rf} with the \texttt{Halofit} recipe \cite{Smith03,Bird:2011rb,Takahashi12}. 

A key aspect of the \texttt{Halofit} prescription employed in this work is its capability to account for the effect of neutrino mass on the non-linear matter power spectrum, thereby correcting the lack of power on the smallest scales (see \cite{Bird:2011rb}). Recently, it has been shown that this revisited version of \texttt{Halofit}, referred sometimes in the literature as ''TakaBird'', remains accurate in the non-linear regime at low redshifts, with deviations below 2\%
for $k\sim 0.3\;h\;\text{Mpc}^{-1}$ at $z=1$ and $\sum m_\nu\sim 0.16\;\text{eV}$ when compared with DEMNUni simulations \cite{Parimbelli:2022pmr}. Therefore, we do not expect significant effects from non-linear corrections to the power spectrum given our specific setup, redshift and
the chosen $k_{\text{max}}$. Furthermore, we must recognize that, even though we model the matter power spectrum in the non-linear regime, this does not make our analysis fully non-linear. A complete treatment would require a consistent modeling of the galaxy bias and redshift-space distortions at those scales.

We have explored three scenarios: a pessimistic case (Pess), with $k_{\text{max}}(\text{GCs}) = 0.10\;h\;\text{Mpc}^{-1}$, an optimistic case (Opt), with $k_{\text{max}}(\text{GCs}) = 0.20\;h\;\text{Mpc}^{-1}$, and an only linear $P(k)$ case, with the maximum scale for the observed power spectrum $k_{\text{max}}(\text{GCs}) = 0.20\;h\;\text{Mpc}^{-1}$. 

\section{Results}

\subsection{Results for J-PAS}

To obtain the sensitivity of galaxy clustering information from J-PAS to $\sum m_\nu$, we performed the forecast by combining the three subsamples with different odds. We compared the results for the optimistic and pessimistic cases, examined the impact of sky coverage and non-linear corrections on the constraints, and performed an analysis using the two tracers, LRG and ELG, both together and separately.

\begin{figure}[H]
\centering
\includegraphics[width=\columnwidth]{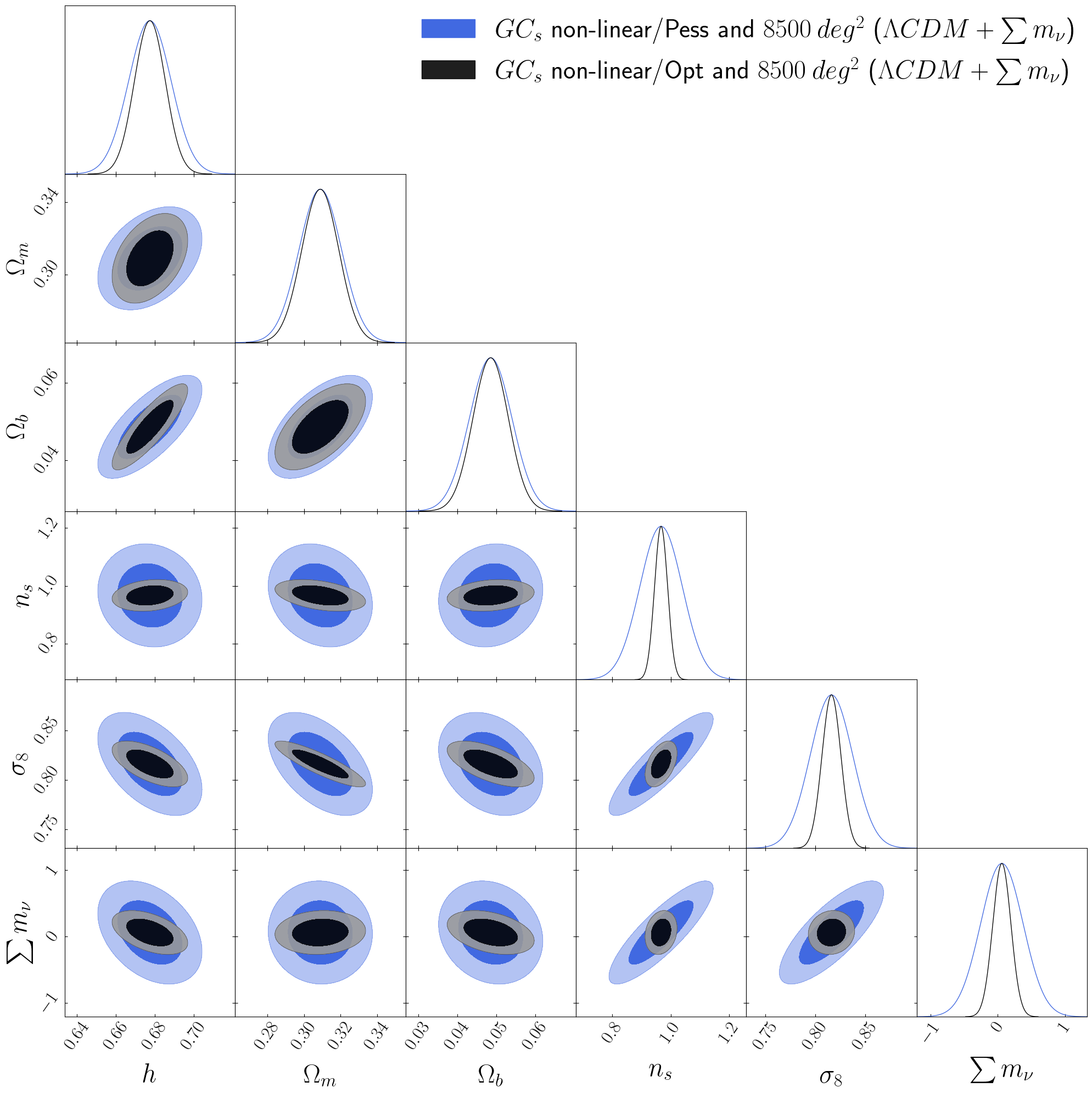}
\caption{Confidence contours at $68\%$ and $95\%$ C.L. for the full survey with  $f_{\text{sky}} = 8500 \:\text{deg}^2$ comparing an Optimistic case with  $k_{\text{max}}(\text{GCs}) = 0.20 \, h \, \text{Mpc}^{-1}$ , and a Pessimistic case with $k_{\text{max}}(\text{GCs}) = 0.10 \, h \, \text{Mpc}^{-1}$, both with non-linear corrections.}
\label{Fig2}
\end{figure}

In Fig.~\ref{Fig2}, we show the comparison between the non-linear pessimistic and optimistic cases in the full survey forecast. A significant improvement in the constraints is observed when smaller scales are considered. This is expected since we are adding more information to the analysis. Especially for neutrinos, the suppression on $P(k)$ becomes increasingly higher at smaller scales, meaning that we have more information about the neutrino mass. 

\begin{figure}[H]
\centering
\includegraphics[width=\columnwidth]{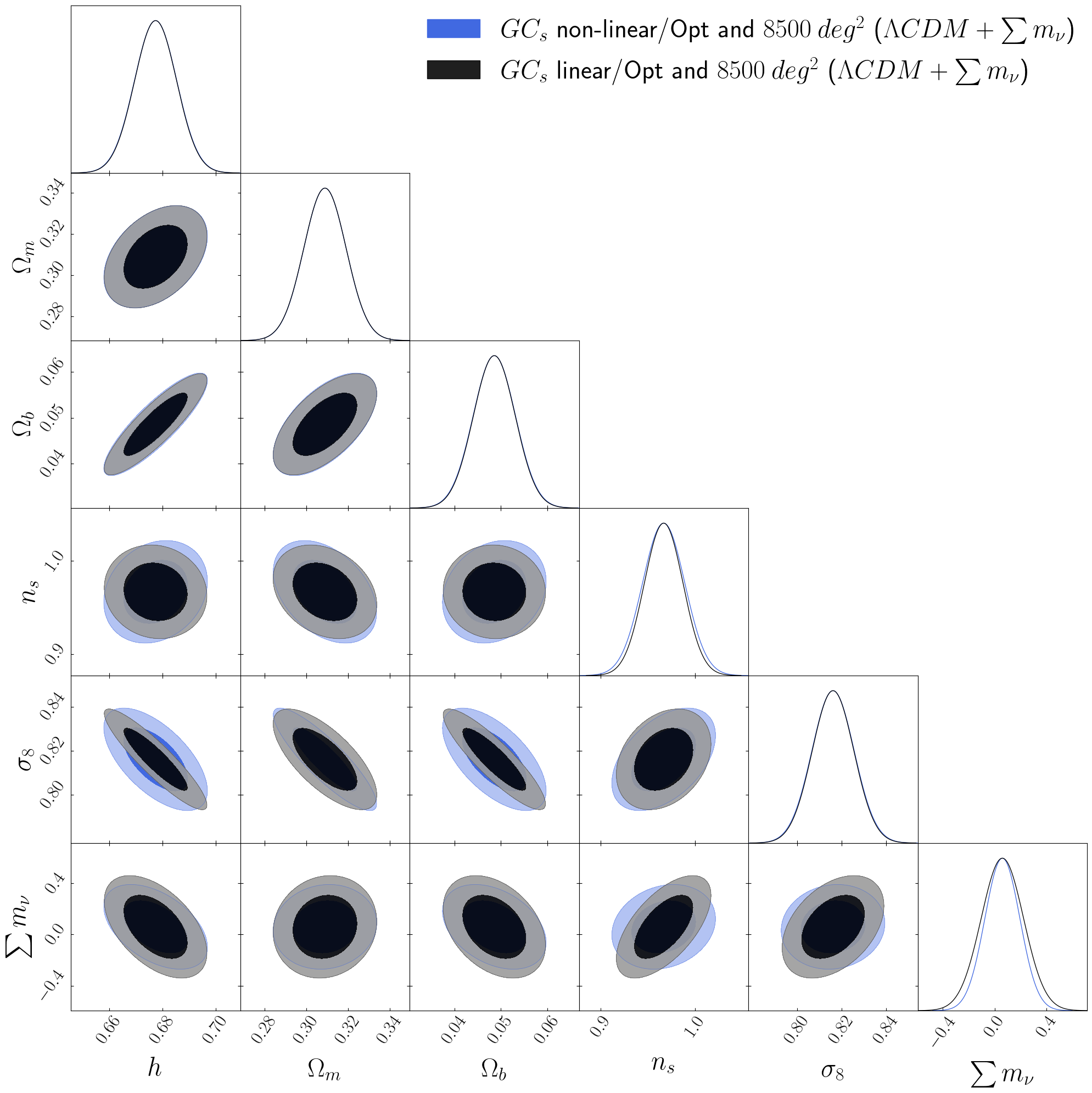}
\caption{Confidence contours at $68\%$ and $95\%$ C.L. for the full survey with $f_{\text{sky}} = 8500 \:\text{deg}^2$ in an Optimistic case with $k_{\text{max}}(\text{GCs}) = 0.20 \, h \, \text{Mpc}^{-1}$, comparing an analysis assuming linear theory to the analysis including non-linear corrections.}
\label{Fig3}
\end{figure}

We also compare the forecast with and without the non-linear corrections in the Opt case, as shown in Fig.~\ref{Fig3}. One can observe that when the non-linear effects are incorporated, they can provide a slight improvement in the constraints on the neutrino mass. However, the magnitude of the improvement depends on the non-linear cutoff scale set by $\Sigma_0$. The cutoff scale $\Sigma_0$ determines the extent of suppression caused by nonlinear effects. A low $\Sigma_0$ means less suppression on non-linear effects, which in turn can improve the precision on neutrino mass constraints. On the other hand, a high $\Sigma_0$ leads to more suppression of non-linear effects, reducing the differences between the linear and non-linear cases and therefore making the non-linear corrections less relevant. Non-linear effects are then most beneficial when low $\Sigma_0$ and larger $k_{\text{max}}(\text{GCs})$ are considered. The value of $\Sigma_0$ is correlated with the amplitude of matter fluctuations $\sigma_8$. For a cosmology with $\sigma_8 \approx 0.80$ the optimal value of $\Sigma_0$ is approximately 11\; $h^{-1}$\; Mpc~\cite{Eisenstein07,Seo07}.

\begin{figure}[H]
\centering
\includegraphics[width=\columnwidth]{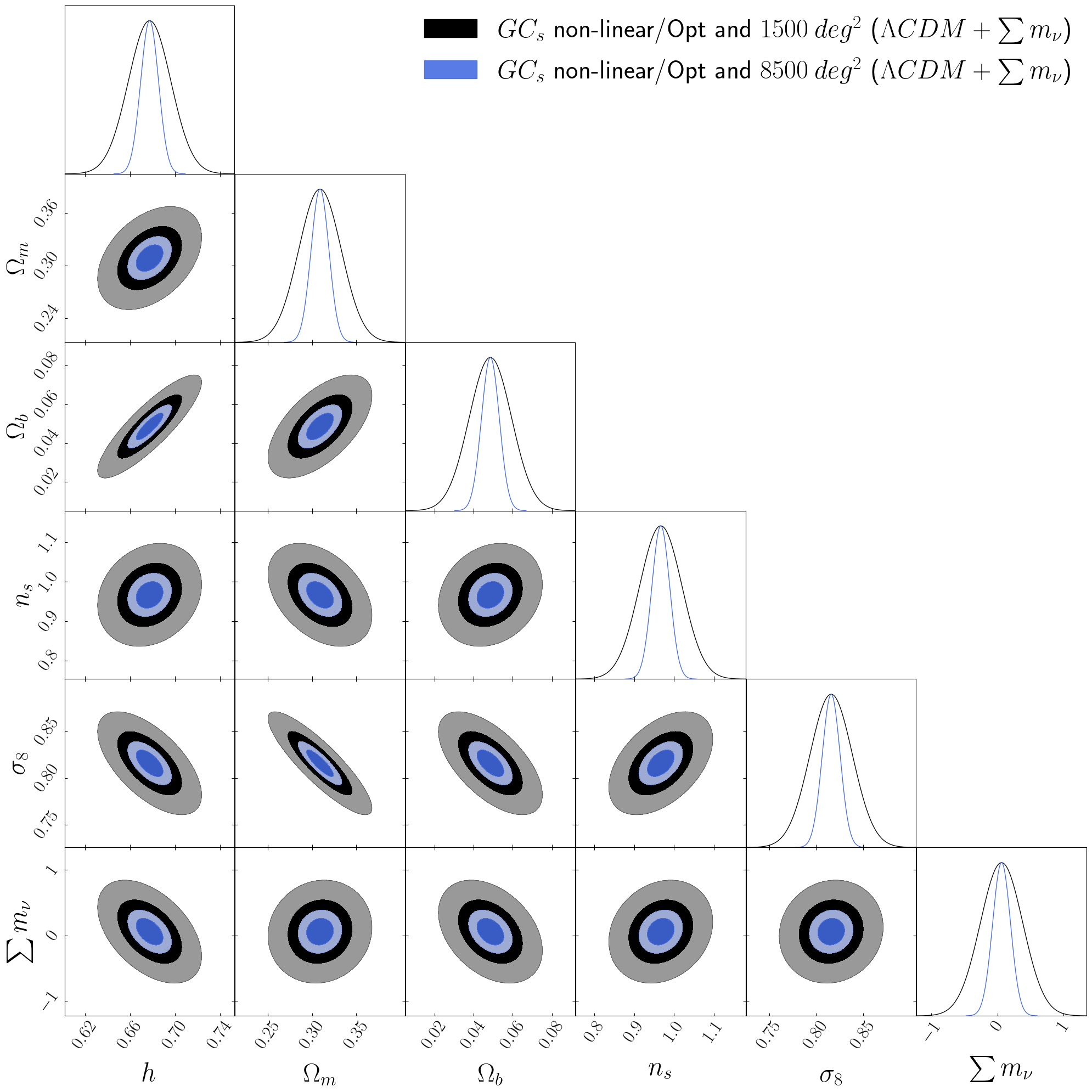}
\caption{Confidence contours at $68\%$ and $95\%$ C.L. comparing a partial survey with $f_{\text{sky}} = 1500 \:\text{deg}^2$ and a full survey with $f_{\text{sky}} = 8500 \:\text{deg}^2$, both for an Optimistic case with $k_{\text{max}}(\text{GCs}) = 0.20 \, h \, \text{Mpc}^{-1}$, and including non-linear corrections.}
\label{Fig4}
\end{figure}

\begin{figure}[H]
\centering
\includegraphics[width=\columnwidth]{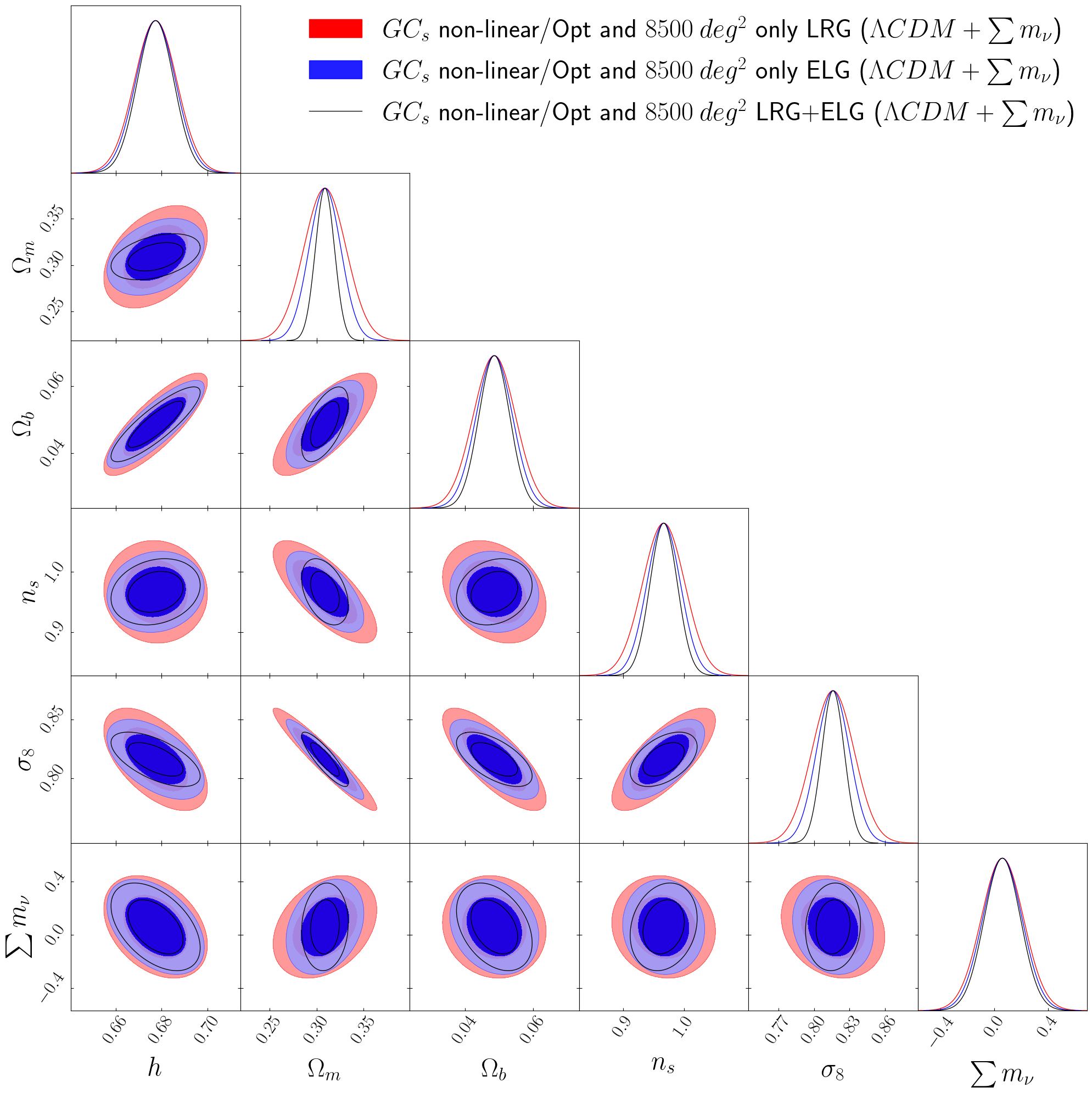}
\caption{Confidence contours at $68\%$ and $95\%$ C.L. for the full survey with $f_{\text{sky}} = 8500 \:\text{deg}^2$ in an Optimistic case with $k_{\text{max}}(\text{GCs}) = 0.20 \, h \, \text{Mpc}^{-1}$, comparing the cosmological constraints from the LRG and ELG samples, and both samples combined.}
\label{Fig5}
\end{figure}

In Fig.~\ref{Fig4}, we show the difference in the results between the two values of the sky coverage explored. We can observe that the power of the constraints scales with the sky coverage as expected. This shows that the sky fraction covered by the survey is an important piece in the constraining power of the experiment.  In Fig.~\ref{Fig5}, we can see the results for each tracer separately, in comparison with the total case (LRG+ELG). We can observe a clear improvement in the constraints provided by the LRG+ELG combination. However, ELGs alone present better constraints than LRGs alone. Hereafter, we will assume that our J-PAS baseline case is: full survey with $f_{\text{sky}} = 8500 \:\text{deg}^2$ and Opt case with $k_{\text{max}}(\text{GCs}) = 0.20 \, h \, \text{Mpc}^{-1}$ with non-linear corrections. 

\begin{figure}[H]
\centering
\includegraphics[width=\columnwidth]{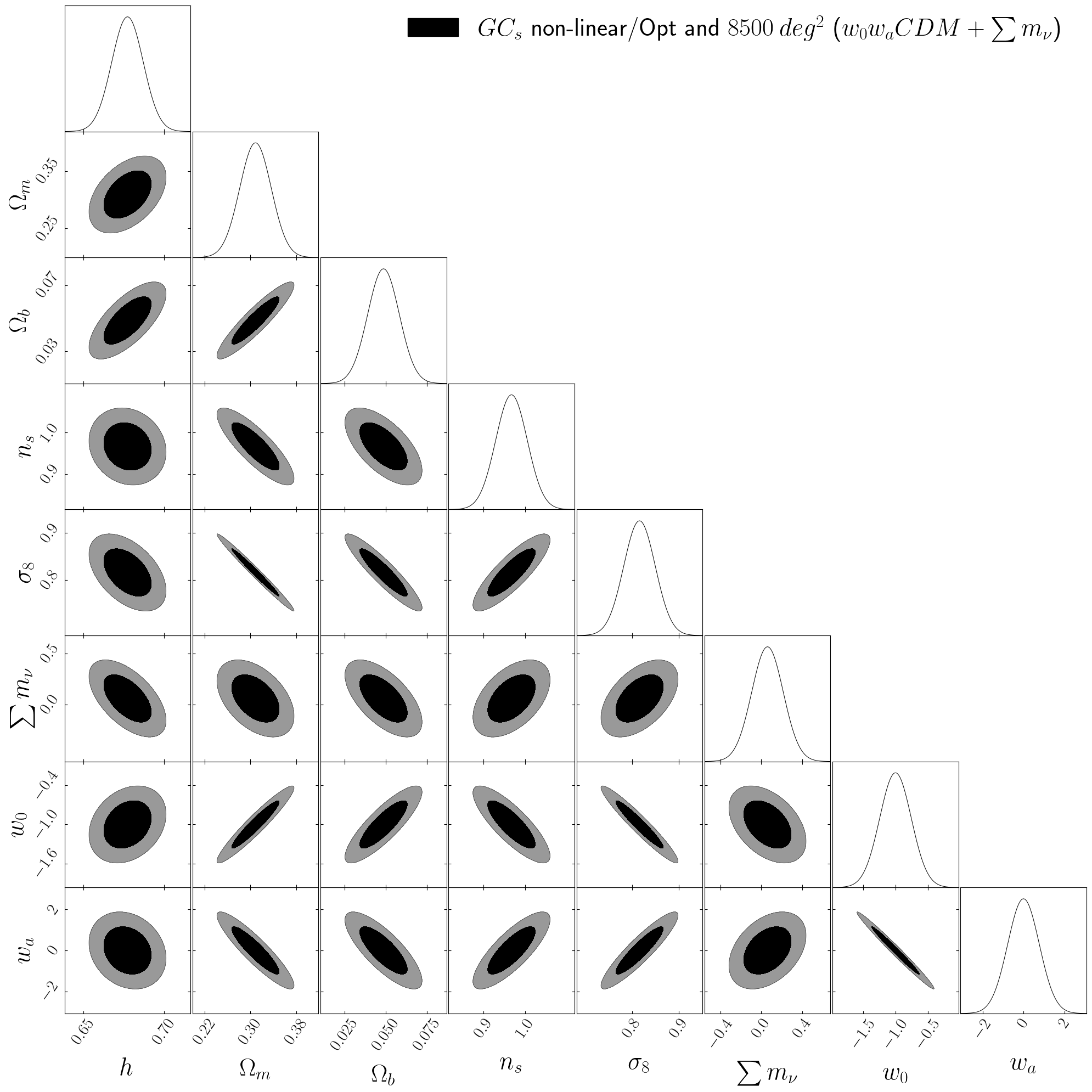}
\caption{Confidence contours at $68\%$ and $95\%$ C.L. for the $w_0w_a CDM + \sum m_\nu$ with full survey with $f_{\text{sky}} = 8500 \:\text{deg}^2$ in an Optimistic case with $k_{\text{max}}(\text{GCs}) = 0.20 \, h \, \text{Mpc}^{-1}$.}
\label{Fig6}
\end{figure}

We show the $w_0w_aCDM + \sum m_\nu$ contours for the J-PAS baseline case in Fig.~\ref{Fig6}. The main thing to note here is that J-PAS $GC_s$ alone will not provide a very tight bound on the dark energy equation of state parameters $w_0$ and $w_a$. We can also observe a moderate correlation between $\sum m_\nu$ and $w_0$ or $w_a$. The orientation of the resulting probability ellipses is in agreement with the results shown by \cite{Archidiacono24} for the EUCLID survey.

In Table~\ref{Tab3}, we present our forecast for the $\Lambda CDM + \sum m_\nu$ and $w_0w_a CDM + \sum m_\nu$ models using the J-PAS baseline case. For neutrino mass, J-PAS $GC_s$ alone will provide us with an upper limit (one-tail) $\sum m_\nu < 0.32$\;eV at 95\%~C.L. for the $\Lambda CDM + \sum m_\nu$ model. This constraint is less tight when considering the $w_0w_a CDM + \sum m_\nu$, with an upper limit of $\sum m_\nu < 0.36$\;eV at 95\% C.L. (for comparison, see \cite{DESI24,Du:2024pai,Shao:2024mag,DESI24FS} for recent constraints on this cosmological model). Given the value of the upper limit for the neutrino mass, this shows that there is no conclusive evidence in favor of normal or inverted ordering for either model with J-PAS $GC_s$ alone, which is consistent with expectations from other future stage IV galaxy surveys. This places our results for the sum of the neutrino mass in a highly competitive position.
\begin{table*}[!ht]
		\centering
		\begin{tabular}{{|l|l|l|}}

			\hline
			\hline
Parameter    & $\Lambda CDM + \sum m_\nu$ & $w_0w_a CDM + \sum m_\nu$ \\  \hline 
            \multicolumn{3}{|c|}{\textbf{J-PAS baseline}}\\
\hline
{\boldmath$h              $} & $ 0.0079         $ & $ 0.0098 $                 \\

{\boldmath$\Omega_m       $} & $ 0.010           $& $0.028           $\\

{\boldmath$\Omega_b       $} & $ 0.0046          $& $0.0095            $\\

{\boldmath$n_s            $} & $0.022              $& $ 0.038          $\\

{\boldmath$\sigma_8       $} & $ 0.0095            $& $ 0.034           $\\

{\boldmath$\sum m_\nu (\mathrm{eV})    $} & $0.26\;(2\sigma)    $& $0.30\;(2\sigma)      $\\

{\boldmath$w_0$} & $-$   & $ 0.24            $\\

{\boldmath$w_a$} & $-$   & $ 0.77               $ \\
\hline
        \end{tabular}
		\caption{Forecasted cosmological constraints at 1$\sigma$ confidence level (2$\sigma$ constraints for the total neutrino mass) for the J-PAS baseline dataset ($GC_s$ non-linear/Opt and $8500\;\text{deg}^2$).  }
        \label{Tab3}
\end{table*}

\subsection{Results for the combination J-PAS+Planck+Pantheon Plus}

To perform the combined analysis, we used Planck 2018 low and high multipole data for temperature, polarization, and lensing ($plikHM+TTTEEE+lowl+lowE+lensing$)~\cite{Planck:2018vyg}, along with Type Ia supernovae data from Pantheon Plus (SN)~\cite{Scolnic_2022,Brout_2022}. Given that the public chains from the Planck collaboration remove negative values for the total neutrino mass (and thus the full probability distribution function could not be recovered), we computed the CMB Fisher matrix directly from the cosmological parameter covariance matrix obtained through the official Planck likelihood. We interfaced the \textit{CLASS} code~\cite{lesgourgues11,Blas:2011rf} with \textit{MontePython}~\cite{Audren:2012wb,Brinckmann:2018cvx} to perform a Monte Carlo Markov Chain (MCMC) numerical analysis. This analysis was applied to the $\Lambda CDM + \sum m_\nu$ and $w_0w_a CDM + \sum m_\nu$ models using Planck 2018 data, both with and without Pantheon Plus data. We then used the covariance matrix from these analyses along with the J-PAS covariance matrix from the Fisher forecast to obtain the combined likelihood function. This function is defined as the product of the probability distribution function from the CMB data covariance matrix (centered at their bestfit values of the cosmological parameters) and the J-PAS Gaussian likelihood (using the Fisher matrix for the J-PAS baseline case), centered at our fiducial cosmology. We note that centering CMB and J-PAS likelihoods at different bestfit values would account for the effect of possible tensions between these two datasets. We emphasize that this combination should be interpreted with caution: when datasets are partially in tension, the apparent tightening of constraints may be influenced by such inconsistencies rather than by genuine complementarity. Therefore, the combined analysis is presented only as an illustration of possible sensitivity improvements, and we do not draw physical conclusions. In this case we used the \texttt{emcee} library to perform MCMC sampling and explore the parameter space from the combined posterior distribution~\cite{Foreman13}. The one-tail 95\% C.L. limits reported hereafter on the sum of neutrino masses $\Sigma m_\nu$ are computed on the truncated posterior distribution, in which the negative values for this parameter have been removed.

\begin{figure}[H]
\centering
\includegraphics[width=\columnwidth]{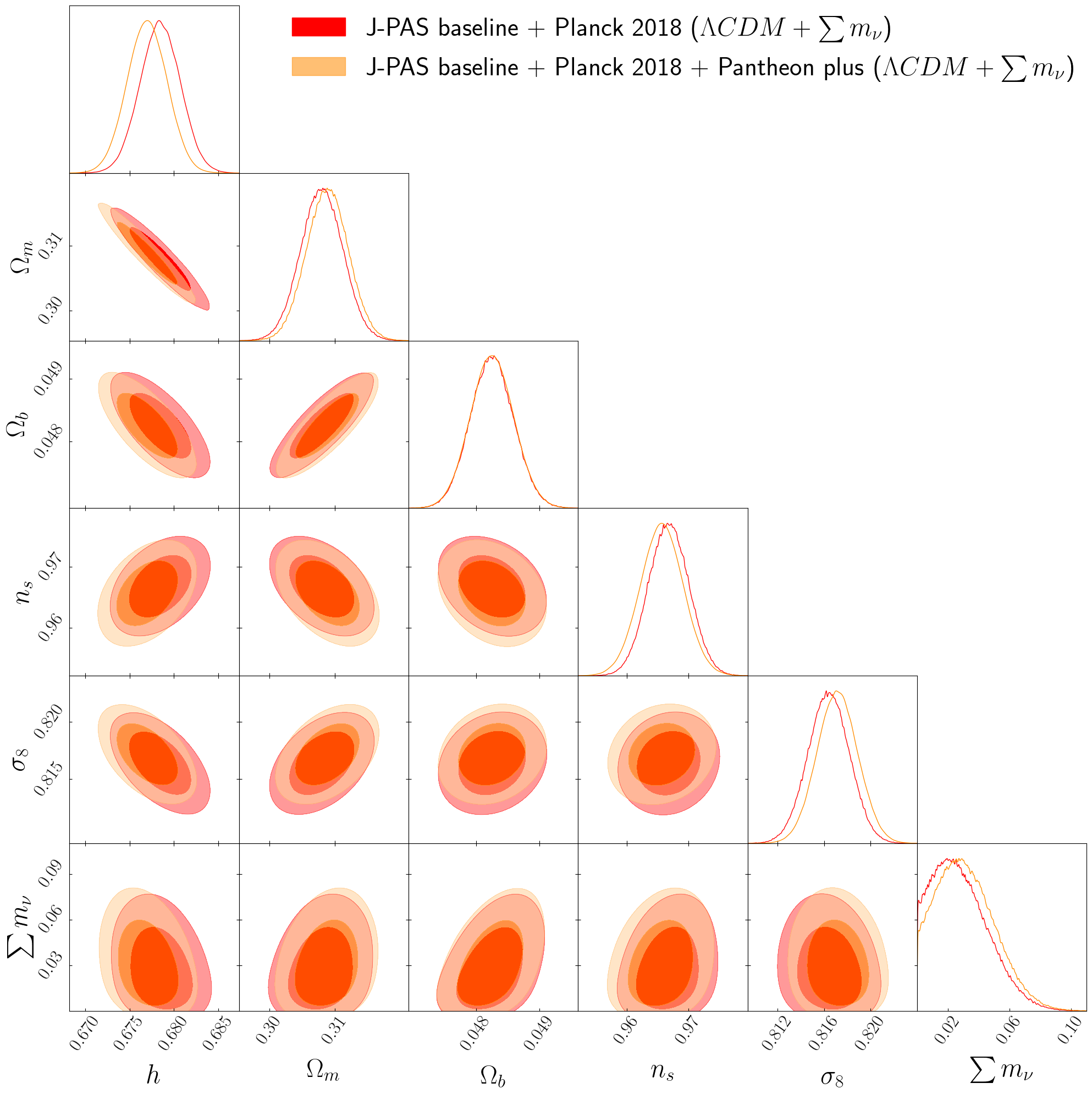}
\caption{Confidence contours at $68\%$ and $95\%$ C.L. for the $\Lambda CDM + \sum m_\nu$ model with the J-PAS forecast with the full survey $f_{\text{sky}} = 8500 \:\text{deg}^2$ in an Optimistic case with  $k_{\text{max}}(\text{GCs}) = 0.20 \, h \, \text{Mpc}^{-1}$ combined with Planck 2018 temperature, polarization and lensing with and without Pantheon Plus SN data.}
\label{Fig8}
\end{figure}

In Fig.~\ref{Fig8}, we present the results for Planck 2018 combined with J-PAS, with and without Pantheon Plus supernovae, for the $\Lambda CDM + \sum m_\nu$ model. We observe that the inclusion of SN makes little difference to the result. In Table~\ref{Tab4}, we present the cosmological constraints obtained for both combinations. We have an upper 95\% one-tail limit of $\sum m_\nu < 0.061$\;eV for the case without supernovae. When we include Pantheon Plus data, this puts our upper limit at $\sum m_\nu < 0.065$\;eV (95\% C.L.) This demonstrates that J-PAS, when combined with Planck, can significantly improve the constraints on neutrino mass.  

\begin{figure}[H]
\centering
\includegraphics[width=\columnwidth]{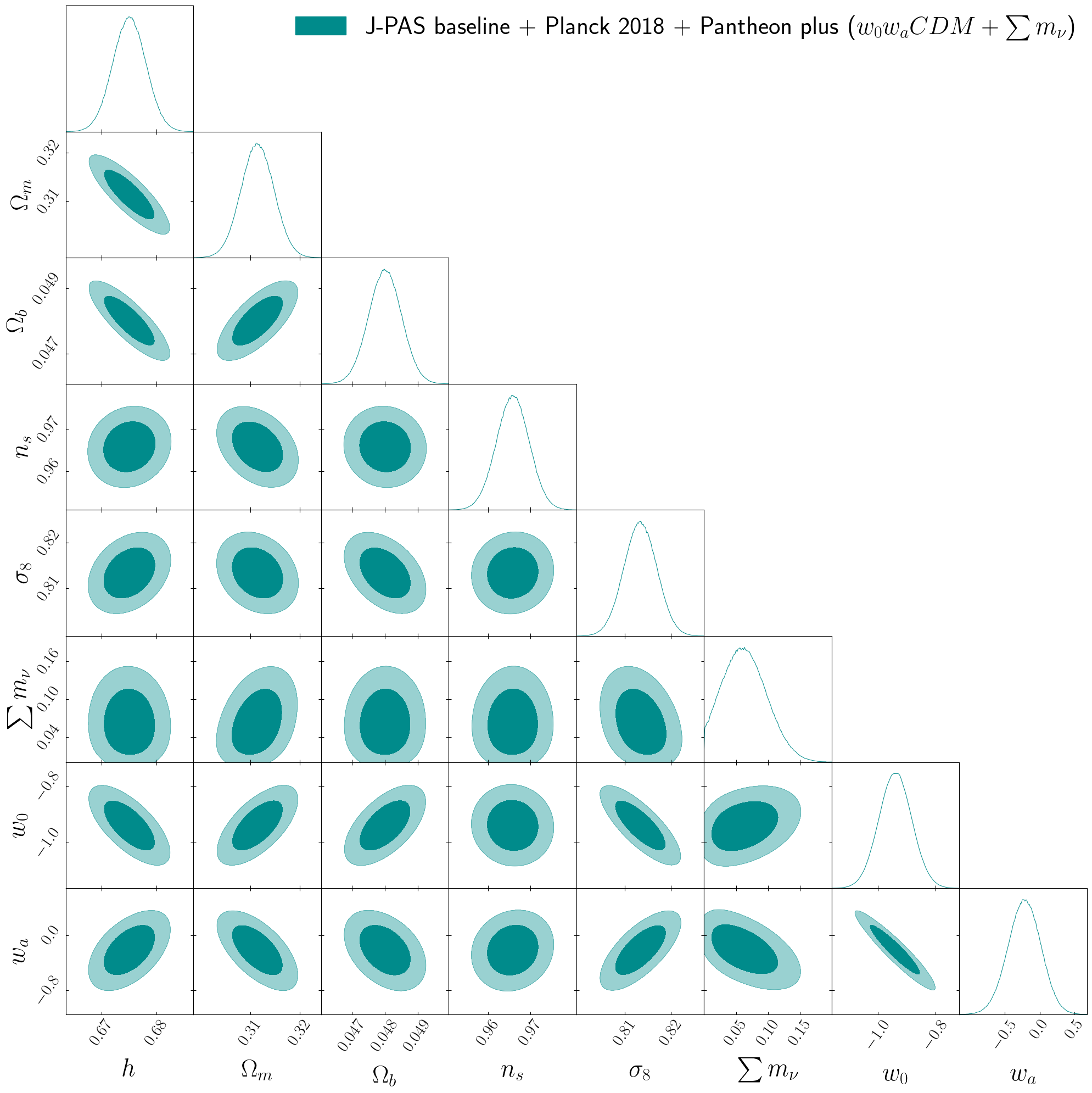}
\caption{Confidence contours at $68\%$ and $95\%$ C.L. for the $w_0w_a CDM + \sum m_\nu$ model with the J-PAS forecast with the full survey $f_{\text{sky}} = 8500 \:\text{deg}^2$ in an Optimistic case with  $k_{\text{max}}(\text{GCs}) = 0.20 \, h \, \text{Mpc}^{-1}$ combined with Planck 2018 temperature, polarization and lensing and Pantheon Plus SN data.  }
\label{Fig10}
\end{figure}

We also performed the analysis for the $w_0w_a CDM + \sum m_\nu$ model. On Fig.~\ref{Fig10}, we show the $w_0w_a CDM + \sum m_\nu$ probability contours for J-PAS combined with Planck 2018 and Pantheon Plus SN data, and on Table~\ref{Tab5} we show the cosmological parameter constraints. When comparing with the $\Lambda CDM + \sum m_\nu$ model, we see a significant loosening of neutrino mass constraints. We obtain an upper limit of $\sum m_\nu < 0.12$\;eV  at 95\% C.L for the combination with J-PAS.

\begin{table*}[!ht]
		\centering
		\begin{tabular}{{|l|l|}}

			\hline
			\hline
Parameter    & $\Lambda CDM + \sum m_\nu$ \\  \hline 
            \multicolumn{2}{|c|}{\textbf{J-PAS baseline + Planck 2018}}\\
\hline
{\boldmath$h              $} & $0.6785\pm 0.0023          $\\

{\boldmath$\Omega_m       $} & $0.3080\pm 0.0032          $\\

{\boldmath$\Omega_b       $} & $0.04826\pm 0.00034        $\\

{\boldmath$n_s            $} & $0.9669\pm 0.0033          $\\

{\boldmath$\sigma_8       $} & $0.8164\pm 0.0018          $\\

{\boldmath$\sum m_\nu (\mathrm{eV})     $} & $0.028^{+0.033}_{-0.028}\;(2\sigma) $  \\  

\hline

            \multicolumn{2}{|c|}{\textbf{J-PAS baseline + Planck 2018 + Pantheon Plus}}\\
\hline

{\boldmath$h              $} & $0.6767\pm 0.0022          $\\

{\boldmath $\Omega_m$} & $0.3093\pm 0.0032          $\\

{\boldmath$\Omega_b       $} & $0.04829\pm 0.00034        $\\

{\boldmath$n_s            $} & $0.9638\pm 0.0035          $\\

{\boldmath$\sigma_8       $} & $0.8169\pm 0.0018          $\\

{\boldmath$\sum m_\nu (\mathrm{eV})$} & $0.032^{+0.033}_{-0.032}\;(2\sigma)   $ \\
     \hline 
        \end{tabular}
		\caption{Forecasted cosmological constraints at 1$\sigma$ confidence level for the $\Lambda CDM + \sum m_\nu$ model, and 2$\sigma$ constraints for the total neutrino mass for J-PAS baseline + Planck (including CMB lensing) and J-PAS baseline + Planck (including CMB lensing) + Pantheon Plus. }
        \label{Tab4}
\end{table*}

\begin{table*}[!ht]
		\centering
		\begin{tabular}{{|l|l|}}

			\hline
			\hline
Parameter    & $w_0w_a CDM + \sum m_\nu$ \\  \hline 

            \multicolumn{2}{|c|}{\textbf{J-PAS baseline + Planck 2018 + Pantheon Plus}}\\
\hline

{\boldmath$h              $} & $0.6750\pm 0.0030          $\\

{\boldmath$\Omega_m       $} & $0.3114\pm 0.0033          $\\

{\boldmath$\Omega_b       $} & $0.04801\pm 0.00050        $\\

{\boldmath$n_s            $} & $0.9659\pm 0.0039          $\\

{\boldmath$\sigma_8       $} & $0.8134\pm 0.0036         $\\

{\boldmath$\sum m_\nu (\mathrm{eV})    $} &  $0.064^{+0.059}_{-0.064}\;(2\sigma)   $ \\

{\boldmath$w_0            $} &  $-0.940\pm 0.057           $\\

{\boldmath$w_a            $} & $-0.22\pm 0.23             $\\

     \hline 
        \end{tabular}
		\caption{Forecasted cosmological constraints at 1$\sigma$ confidence level for the $w_0w_a CDM + \sum m_\nu$ model, and 2$\sigma$ constraints for the total neutrino mass for J-PAS baseline + Planck (including CMB lensing) and J-PAS baseline + Planck (including CMB lensing) + Pantheon Plus. }
        \label{Tab5}
\end{table*}

\begin{figure}[H]
\centering
\includegraphics[width=\columnwidth]{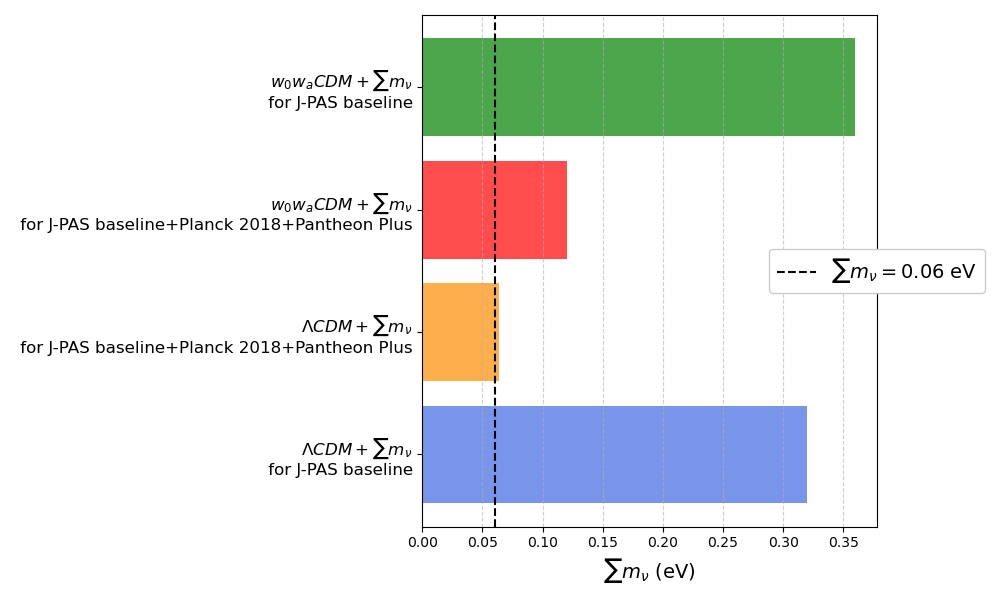}
\caption{Neutrino mass upper limits for $\Lambda CDM + \sum m_\nu$ and $w_0w_a CDM$. The green and blue bars represent J-PAS alone for $w_0w_a CDM + \sum m_\nu $ and $\Lambda CDM + \sum m_\nu$, respectively. The red and yellow bars is J-PAS combined with CMB and SN. The black dashed line indicates the normal hierarchy limit for the neutrino mass.}
\label{Fig11}
\end{figure}

It is worth noting that on all the combined analyses where J-PAS is combined with external data, we have excluded the non-physical regions for the sum of the neutrino masses, i.e., regions for $\sum m_\nu < 0$. The results show that the excluded regions are not a significant part of the constraint, indicating that any preference for a negative neutrino mass is thinned for this data configuration.  Therefore, J-PAS, when combined with external data, mainly CMB, will achieve excellent sensitivity in measuring the sum of the neutrino masses, comparable to the measurements expected from stage IV surveys. In Fig.~\ref{Fig11}, we present a summary of our predicted results for J-PAS alone and in combination with external data. Our constraints on the total neutrino mass are competitive for both the $\Lambda CDM + \sum m_\nu$ and $w_0w_a CDM + \sum m_\nu$ models when compared with the latest results reported by the DESI Collaboration~\cite{DESI25Neutrino}.


\section{Discussion and conclusion}

In this paper we have presented a Fisher matrix forecast on neutrino mass constraints from J-PAS galaxy power spectrum measurements and their combination with CMB measurements from the Planck Collaboration and SN from Pantheon Plus. We conducted a series of tests to see how sensitive our forecasted constraints are to different choices for the survey settings as well as for the cosmological analysis, before including the information from cosmic microwave background and type Ia supernovae. Besides the trivial scaling with survey area (comparing, for instance, a 1,500 deg${^2}$ versus a 8,500 deg${^2}$ survey), one can choose to include clustering data at small scales, relying on the modeling of non-linear scales given by \texttt{Halofit}. Increasing the maximum wavenumber of the power spectrum from a pessimistic value of 0.10~$h\,\mathrm{Mpc}^{-1}$ to an optimistic value of 0.20~$h\,\mathrm{Mpc}^{-1}$ significantly improves the survey's constraining power. The EUCLID collaboration finds only a small difference when extending the maximum scale from 0.25~$h\,\mathrm{Mpc}^{-1}$ to 0.30~$h\,\mathrm{Mpc}^{-1}$, except for theoretical errors and systematic effects that may have been overlooked in the modeling of the observables (see \cite{Archidiacono24}). Similarly, applying a linear theory treatment of clustering at such small scales results in only a minor difference that can be safely neglected, although other cosmological parameters, such as $n_s$ and $\sigma_8$, are more sensitive to this choice. This shows that beyond a certain scale, the incremental gain in constraining power becomes marginal, particularly when theoretical uncertainties are taken into account. Interestingly, splitting the galaxy sample by type reveals that neutrino mass constraints from emission line galaxies (ELGs), which have a higher number density, are stronger than those from luminous red galaxies (LRGs). In the multitracer case using both LRGs and ELGs, the improvement in the total neutrino mass bounds is negligible, but the inclusion of LRGs still enhances constraints on other cosmological parameters. This makes ELGs particularly valuable tracers for obtaining competitive constraints on the sum of neutrino masses.

We have shown that an 8,500 sq. deg. survey from J-PAS, given the galaxy number density over the redshift range covered, can place competitive neutrino mass constraints. The constraints from density fluctuations as extracted by galaxy clustering information from the galaxy power spectrum break the degeneracies present in background measurements such as those traced by baryon acoustic oscillations or the cosmic microwave background (which is itself mostly a background measurement, \cite{Bertolez-Martinez:2024wez}). In particular, we have combined the clustering Fisher matrix from J-PAS with the actual measurements from Planck temperature, polarization and lensing power spectra at low and high multipoles, and Type Ia Supernovae from Pantheon Plus. We have also tested the numerical stability of the resulting Fisher matrix by checking the sensitivity of the numerical derivatives used when computing the clustering Fisher matrix. If J-PAS future data shows to be consistent with the assumed fiducial parameter values and forecast uncertainties as reported in this work, the combination with Planck for the $\Lambda CDM + \sum m_\nu$ model would return an upper limit of $\Sigma m_\nu < 0.061$\;eV (95\% C.L.). Recently, other works have shown tighter constraints; however, those use a larger number of cosmological probes than this forecast. Although it is well known that EUCLID will have a sensitivity of $0.02$\;eV when combined with Planck data \cite{EUCLID:2011zbd, Audren:2012vy, Sprenger:2018tdb,Archidiacono24}, those forecasts assume a combination of clustering and lensing information (Euclid primary probes) with Planck measurements, while in our baseline dataset we did not include J-PAS galaxy lensing information to separate more cleanly large-scale structure from cosmic microwave background datasets, since the latter includes CMB lensing. In this sense, our constraints are conservative and, if J-PAS cosmic shear measurements are included, they can result in further improvement of these cosmological constraints. In order to compare on equal footing the forecasts from J-PAS and EUCLID for the $\Lambda CDM$+$\sum m_\nu$ model, we should compare the results displayed in Table \ref{Tab3} using J-PAS baseline, with the ones provided in Table 5 of \cite{Archidiacono24} using galaxy clustering (GC$_{\text{sp}}$) data. The error bars estimated with J-PAS (EUCLID) in each case are $\delta h = 0.0079$ (0.033), $\delta \Omega_m = 0.010$ (0.0068), $\delta \Omega_b = 0.0046$ (0.0037), $\delta n_s = 0.022$ (0.029) and $\delta \sigma_8 = 0.0095$ (0.0077). Therefore, for the $h$ and $n_s$ parameters with J-PAS baseline we obtain error bars 76\% and 20\% smaller than with EUCLID GC$_{\text{sp}}$ data, whereas for the parameters $\Omega_m$, $\Omega_b$ and $\sigma_8$ EUCLID gets respectively, 31\%, 20\% and 19\% smaller error bars. As for the sum of the neutrino mass parameter, both J-PAS and EUCLID GC$_{\text{sp}}$ forecast a 95\% upper bound of 0.32 eV on $\sum m_\nu$.  

We also performed this analysis for the $w_0w_a CDM + \sum m_\nu$ model, where we found a relaxation of the neutrino mass constraints. Specifically, combining J-PAS with CMB and SN data, we obtained an upper bound of $\sum m_\nu < 0.12$\;eV (95\% C.L.). This clearly demonstrates that the cosmological model will play a central role in future discussions about a potential tension between neutrino mass measurements from cosmological observables and neutrino oscillation experiments. Additionally, the J-PAS dataset will allow us to explore different models that modify the neutrino sector, potentially resolving important issues in the standard model of cosmology, such as the Hubble tension.

Our results show that the constraints expected from J-PAS will be comparable to those forecasted for other forthcoming large-scale structure surveys, such as Euclid and LSST, and will provide a valuable cross-check for space-based clustering constraints from other Stage IV galaxy surveys. Our constraints assume a neutrino mass best-fit value from J-PAS equal to the minimum allowed by neutrino oscillation data assuming normal hierarchy mass ordering, which is about 0.06\;eV. Recent claims, especially since the DESI survey first year results \cite{Craig24,Green24,Elbers24,Naredo-Tuero:2024sgf,Jiang:2024viw}, have shown a preference for zero or negative neutrino mass which might signal an inconsistency in the assumed cosmological model. However, the results from this forecast show that the negative neutrino mass will show tension with neutrino oscillation measurements and that J-PAS will provide a useful cosmological dataset to evaluate the significance of these issues. These results show the forecasted constraining power of the clustering measurements from the J-PAS survey, which in the case of the neutrino mass data are particularly promising. The need of independent datasets that confirm or rule out claims from other large galaxy surveys such as DESI, EUCLID, or LSST, makes the J-PAS dataset essential to address the forthcoming challenges that the current cosmological model is facing.


\section*{Acknowledgements}
GR is supported by the Coordena\c{c}\~ao de Aperfei\c{c}oamento de Pessoal de N\'ivel Superior (CAPES). AJC acknowledges support from the Spanish Ministry of Science, Innovation, and Universities (PID2022-140440NB-C21).
This research was financially supported by the project "Plan Complementario de I+D+i en el \'area de Astrof{\'\i}sica" funded by the European Union within the framework of the Recovery, Transformation and Resilience Plan - NextGenerationEU and by the Regional Government of Andaluc{\'i}a (Reference AST22\_00001). AJC thanks the hospitality of Departamento de F{\'\i}sica Te\'orica y del Cosmos at Universidad de Granada. JSA is supported by CNPq grant No. 307683/2022-2 and Funda\c{c}\~ao de Amparo \`a Pesquisa do Estado do Rio de Janeiro (FAPERJ) grant No. 259610 (2021). ALM is supported by the MICIN (Spain) Project No. PID2022-138263NB-I00 funded by MICIU/AEI/10.13039/501100011033 and by ERDF/EU. MM has been supported by the Spanish Ministry of Science, Innovation and Universities MICIU/AEI/10.13039/501100011033/ (grant PID2022-14044NB-C21) and by Junta de Andaluc{\'\i}a (grant FQM 101). JGR acknowledges financial support from the Programa de Capacita\c{c}\~ao Institucional do Observat\'orio Nacional (PCI/ON/MCTI) and the Funda\c{c}\~ao de Amparo \`a Pesquisa do Estado do Rio de Janeiro (FAPERJ) grant No. E-26/200.513/2025. FBMS is supported by Conselho Nacional de Desenvolvimento Científico e Tecnológico (CNPq) grant No. 151554/2024-2. C.H.-M. acknowledges the support of the Spanish Ministry of Science and Innovation project PID2021-126616NB-I00. 

Based on observations made with the JST/T250 telescope and JPCam at the Observatorio Astrofísico de Javalambre (OAJ), in Teruel, owned, managed, and operated by the Centro de Estudios de Física del Cosmos de Aragón (CEFCA). We acknowledge the OAJ Data Processing and Archiving Unit (UPAD) for reducing and calibrating the OAJ data used in this work.
Funding for the J-PAS Project has been provided by the Governments of Spain and Aragón through the Fondo de Inversión de Teruel, European FEDER funding and the Spanish Ministry of Science, Innovation and Universities, and by the Brazilian agencies FINEP, FAPESP, FAPERJ and by the National Observatory of Brazil. Additional funding was also provided by the Tartu Observatory and by the J-PAS Chinese Astronomical Consortium. This paper has gone through internal review by the J-PAS collaboration.



\bibliography{bibliography}

\end{document}